\documentclass[12pt]{article}
\usepackage{graphicx}

\advance\textwidth by 3truecm
\advance\oddsidemargin by -1.5truecm

\newcommand\HF{{\mathrm{HF}}}

\begin{document}
\title{\normalsize\bf
Bond-order correlation energies for small Si-containing molecules
   compared with \emph{ab initio} results from low-order
   M\o{}ller-Plesset perturbation theory} 
\author{\normalsize A. Grassi, G. M. Lombardo, G. Forte\\
\normalsize\emph{Dipartimento di Scienze Chimiche, Facolt\`a di Farmacia,
   Universit\`a di Catania,}\\ 
\normalsize\emph{Viale A. Doria, 6, I-95126 Catania, Italy}\\[\baselineskip]
\normalsize G. G. N. Angilella, R. Pucci\\
\normalsize\emph{Dipartimento di Fisica e Astronomia, Universit\`a di
   Catania, and}\\
\normalsize\emph{Istituto Nazionale per la Fisica della Materia, UdR Catania,}\\
\normalsize\emph{Via S. Sofia, 64, I-95123 Catania, Italy}\\[\baselineskip]
\normalsize N. H. March\\
\normalsize\emph{Department of Physics, University of Antwerp,}\\
\normalsize\emph{Groenenborgerlaan 171, B-2020 Antwerp, Belgium, and}\\
\normalsize\emph{Oxford University, Oxford, UK}}

\date{}

\maketitle

\abstract{%
The present study of small molecules containing silicon has been
   motivated by (a) the considerable interest being shown currently in
   the kinetics and reactivity of such molecules, and (b) the
   biotechnological potential of silicon-derivate surfaces as
   substrates in the adsorption of, for instance, amino acids and
   proteins.
Therefore, we have studied by (i) a semi-empirical approach and (ii)
   an \emph{ab initio} procedure employing low-order M\o{}ller-Plesset
   perturbation theory, the molecular correlation energies of some
   neutral closed and open shell silicon-containing molecules in the
   series SiX$_n$Y$_m$.
Procedure (i) is shown to have particular merit for the correlation of
   the ionic members studied in the above series, while the \emph{ab
   initio} procedures employed come into their own for neutral
   species. 
}

\section{Introduction}

In the last twenty years, much work has been carried out in order to
   calculate the correlation energy in atoms and molecules.
In particular, in atoms, March and Wind \cite{March:92a} showed how it
   is possible to recover the main trend of the semi-empirical
   correlation energies obtained long ago by Clementi
   \cite{Clementi:63}, using simple arguments based on the correlation
   energy density (see also \cite{Alonso:03}).

Within the framework of Density Functional Theory (DFT), many studies
   have been made to develop new correlation energy functionals.
These are now available as a result of the work of Vosko, Wilk and
   Nusair (VWN) \cite{Vosko:80}, Lee, Yang and Parr (LYP)
   \cite{Lee:88}, and Perdew and Wang (PW) \cite{Perdew:92}, and have
   all been introduced into some  quantum chemistry packages
   ({\sc gaussian}, {\sc nwchem}, {\sc gamess}, etc).

In a previous study (hereafter denoted as Paper~I) \cite{Grassi:96},
   some of the present authors have introduced a new
   semi-empirical technique in order to calculate the electron
   correlation energy in molecules.
Essential to this work was the division of the total electron density
   of the molecule into parts belonging to specific nuclei, and a part
   belonging solely to the chemical bond and therefore, by using
   semi-empirical modelling,  strictly correlated with the molecular
   bond-order (BO).
One important conclusion of this work was that the molecular
   correlation energy is generally larger, in absolute value, than the
   sum of the correlation energies of the separated atoms.
It is known from the pioneering work of Pauling, Mulliken, Coulson and
   other authors \cite{Coulson:39,Pauling:35,Mulliken:59}, that bond
   order and molecular bond length are strongly correlated.
In this sense our work is somewhat related to the work of Fulde
   \emph{et al.} \cite{Pfirsh:85,Oles:86}, where an analytical model
   of  molecular correlation energies based on bond lengths was
   proposed. 

In Paper~I we developed a parameterization for compounds containing
   only hydrogen, oxygen and carbon atoms and the related bonds X--Y
   (X, Y = C, H, O).
The very good results obtained in the calculation of the correlation
   energy for some molecules containing these elements, compared with
   the experimental values, encouraged us to extend this approach to
   second-row elements and in particular to silicon-containing
   compounds, like fluorine and chlorine-substituted silylenes, silyl
   radicals and silanes.
It is relevant in this context to note that Schlegel \emph{et al.}
   \cite{Schlegel:84,Ignacio:90,Su:93} have calculated the heats of
   formation of these compounds by \emph{ab initio} methods,
   introducing part of the correlation energy using the standard
   M\o{}ller-Plesset perturbative expansion technique.

We stress at this point that the kinetics and reactivity of small
   silicon-containing molecules are attracting considerable interest,
   due to the quite different reaction mechanism of the silicon
   compounds with respect to the carbon compounds.
For example, pyrolysis of silanes yields silylenes rather than silyl
   radicals, whereas photolysis produces silyl radicals
   \cite{Gaspar:81}.
Moreover, silylenes and silyl radicals with various degrees of
   fluorination (or chlorination) play an important role in chemical
   vapor deposition \cite{Scott:81,Haller:83,Robertson:84}.
But the main area of interest of these compounds are the
   silicon-derivate surfaces which are widely used as substrates in
   the adsorption process of biological compounds, such as
   amino~acids, proteins etc, in view of their application in
   different biotechnology areas such as biomaterials, biosensors, and
   bioseparation \cite{Gambino:04}.
 
In this study we have investigated the correlation energies of
   SiX$_n$Y$_m$ (X, Y = H, F, Cl; $n+m=4$, 3, 2) compounds using both
   standard \emph{ab initio} techniques, at various levels of
   approximation, and the bond-order method developed in Paper~I.
In section~\ref{sec:methods} we describe briefly the theoretical
   methods used to estimate the molecular correlation energy.
M\o{}ller-Plesset theoretical techniques are summarized in
   subsection~\ref{ssec:abinitio} while the Bond-Order Correlation
   Energy (BOCE) method is described in subsection~\ref{ssec:boce}.
Section~\ref{sec:results} is concerned with the calculated correlation
   energies of SiX$_n$Y$_m$ compounds and the ionization potentials
   (IP) of some of these molecules together with a comparison with the
   experimental values.
Our final conclusions are recorded in Section~\ref{sec:summary}. 

\section{Methods}
\label{sec:methods}

\subsection{\emph{Ab initio} methods}
\label{ssec:abinitio}

\emph{Ab initio} molecular orbital Hartree-Fock calculations (HF) were
   performed with the {\sc gaussian~03} package \cite{Frisch:03} using
   the polarization 6-31G$^{\ast\ast}$ basis sets.
The 6-31G$^{\ast\ast}$ basis contains a set of six $d$ Cartesian
   functions on each of the heavy atoms, and a set of three $p$
   function on each hydrogen.
Equilibrium geometries were obtained by full optimization using
   analytical gradient methods \cite{Schlegel:82}.
Electron correlation energy was estimated with M\o{}ller-Plesset (MP)
   perturbation theory \cite{Moeller:34} up to fourth order (MP4),
   including all electrons and considering single, double excitations.
Full molecular geometry optimization was performed up to third order
   M\o{}ller-Plesset level (MP3).
Quadruple excitations (MP4SQDT) were introduced by fixing the
   molecular geometry obtained from MP3 calculations.
For both HF and MP calculations, restricted Hartree-Fock (RHF) and
   unrestricted Hartree-Fock (UHF) methods were applied to closed and
   open shell systems, respectively.

We stress, at the outset, that the emphasis of this work is the
   correlation energy.
To present such results on a large number of Si-containing compounds,
   we have deliberately restricted the basis sets used.
To extend these is straightforward for the future, should it be deemed
   instructive to do so.

\subsection{Bond-Order Correlation Energy (BOCE) Method}
\label{ssec:boce}

The `experimental' correlation energy $E_c^{\exp}$ is defined as
\begin{equation}
E_c^{\exp} = E_S - E_\HF,
\label{eq:S-HF}
\end{equation}
where $E_\HF$ denotes the (molecular) Hartree-Fock energy and $E_S$ is
   the so-called Schr\"odin\-ger (or ``true'') energy.
In the case of atoms, the latter is obtained from the sum of the
   experimental ionization energies, plus Bethe-Salpeter and
   relativistic corrections.
In the case of molecules, $E_S$ also includes smaller contributions,
   taking into account for the other molecular degrees of freedom, such
   as vibrational and rotational motions.
In both cases, $E_S$ can be constructed from available data
   \cite{Lide:94} following the standard procedure described in
   Cremer's papers \cite{Cremer:82,Cremer:82a}.
By using this procedure and the available experimental data, we have
   obtained $E_S$ for all molecules in the series SiX$_n$Y$_m$ (X, Y =
   H, F, Cl; $n+m=4$, 3, 2). 
Although $E_c^{\exp}$ in Eq.~(\ref{eq:S-HF}) is usually referred to as
   an `experimental' correlation energy, it obviously contains some
   theoretical input through the Hartree-Fock energy $E_\HF$, whose
   degree of approximation can be controlled by choosing a
   sufficiently large basis set.
Moreover, for some molecules, whenever the experimental data for the
   heat of formation, or the vibrational molecular frequencies were
   unavailable, we have made recourse to available theoretical
   estimates of these quantities, as indicated in the table captions.
Fig.~\ref{fig:schema} summarizes the definitions of the experimental
   correlation energies, as well as of other various energies
   addressed in turn below.

Within the bond-order approach \cite{Grassi:96}, the correlation
   energy can be estimated as follows.
Given the formation reaction of a generic molecule,
\begin{equation}
A + B \longrightarrow C + \mbox{binding energy},
\end{equation}
the molecular energy can be partitioned as
\begin{equation}
E(C) = E(A) + E(B) - \mbox{binding energy} = E(A) + E(B) + E(AB),
\label{eq:binding}
\end{equation} 
where $E(AB)$ is the energy due to the $A$--$B$ bond formation.
Then, in general, the Schr\"odinger energy $E_S$ of a polyatomic
   molecule can be written as
\begin{equation}
E_S = \sum_A E_S (A) + \sum_{{\mbox{\scriptsize all $AB$}}} E_S (AB).
\label{eq:S}
\end{equation}
Likewise, in the HF approximation, the total molecular energy can be
   expressed in the form
\begin{equation}
E_\HF = \sum_A E_\HF (A) + \sum_{\mbox{\scriptsize all $AB$}} E_\HF (AB).
\label{eq:HF}
\end{equation}
Subtracting Eq.~(\ref{eq:HF}) from Eq.~(\ref{eq:S}), one obtains the
   theoretical correlation energy as
\begin{eqnarray}
E_c^{\mathrm{theo}} &=& \sum_A [ E_S (A) - E_\HF (A)] +
   \sum_{\mbox{\scriptsize all $AB$}} [ E_S (AB) - E_\HF (AB) ] \nonumber\\
&=& \sum_A E_c^{\mathrm{exp}} (A_{\mbox{\scriptsize in molecule}} ) +
   \sum_{\mbox{\scriptsize all $AB$}} E_c^{\mathrm{theo}} (AB),
\label{eq:theo}
\end{eqnarray}
where we take as known the Schr\"odinger atomic energy from
   experimental data, as explained in the next section. 
In Eq.~(\ref{eq:theo}), $E_c^{\mathrm{exp}} (A_{\mbox{\scriptsize in
   molecule}} )$ 
   is the effective correlation energy for the atom $A$  \emph{in the
   molecule.}
This quantity is obtained from $E_c^{\mathrm{exp}} (A)$ relative to an
   isolated atom $A$ and rescaled taking into account the effective
   number of non-bonding electrons of the atom $A$ in the molecule.
In particular, if $n$ is the total number of electrons of $A$ involved
   in the bonds and $Z$ its atomic number, $E_c^{\mathrm{exp}}
   (A_{\mbox{\scriptsize in molecule}} )$ is obtained as
\begin{equation}
E_c^{\mathrm{exp}} (A_{\mbox{\scriptsize in molecule}} ) =
   E_c^{\mathrm{exp}} (A) \frac{Z-n}{Z} .
\label{eq:Zn}
\end{equation}

The next step in the present approximation scheme is to consider
   $E_c^{\mathrm{theo}} (AB)$ as an analytical function of the
   bond-order $P_{AB}$ between the two atoms $A$ and $B$ of the
   molecule.
In practice, we have assumed a linear form for $E_c^{\mathrm{theo}}
   (AB)$ as a function of $P_{AB}$, namely
\begin{equation}
E_c^{\mathrm{theo}} (AB) = a_{AB}  P_{AB} ,
\label{eq:Loewdin}
\end{equation}
where $a_{AB}$ is a parameter which depends on the $A$--$B$ bond.
Moreover, the  $P_{AB}$ bond-order, according to L\"owdin's definition
   \cite{Loewdin:50}, can be written as
\begin{equation}
P_{AB} = \sum_\mu^A \sum_\nu^B D_{\mu\nu} D_{\mu\nu} ,
\end{equation}
with $D= S^{1/2} T S^{1/2}$, $S$ being the overlap matrix and $T$ the
   first-order density matrix obtained from the HF calculation.
One then obtains:
\begin{equation}
E_c^{\mathrm{theo}} = \sum_A E_c^{\mathrm{exp}}
   (A_{\mbox{\scriptsize in molecule}} ) + \sum_{\mbox{\scriptsize all
   $AB$}} a_{AB} P_{AB} .
\end{equation}
The parameters $a_{AB}$ can be obtained from a model system which
   contains the $A$--$B$ bond and for which is it possible to
   calculate the Schr\"odinger molecular energy $E_S$ and then the
   experimental correlation energy $E_c^{\mathrm{exp}}$.
After obtaining the best values of the $a_{AB}$ parameters for each
   $A$--$B$ bond, application of Eq.~(\ref{eq:Loewdin}) then
   permits to calculate the theoretical correlation energy
   $E_c^{\mathrm{theo}}$.
In Appendix~\ref{app:bond} we report a specific example to illustrate the
   procedure used to obtain the $a_{AB}$ parameters.

\section{Results and discussion}
\label{sec:results}

\subsection{Neutral closed and open shell SiX$_n$Y$_m$ compounds}
\label{ssec:neutral}

In order to calculate the correlation energy of fluorine and
   chlorine-substituted silylenes compounds, we have obtained the bond
   parameter $a_{AB}$ of Eq.~(\ref{eq:Loewdin}) using the experimental
   correlation energy $E_c^{\mathrm{exp}}$ of some model molecules and
   the experimental atomic correlation energies. 
In Tab.~\ref{tab:HF} we report the Schr\"odinger and the HF atomic
   energies of the constituent atoms in their fundamental state and
   the related correlation energies.
In Tab.~\ref{tab:bond} we record results for some model systems
   together with the bond parameter extracted from
   $E_c^{\mathrm{exp}}$.
All energy values are in Hartrees.

In the second column of Tab.~\ref{tab:theo} we record the
   Schr\"odinger energy of the SiX$_n$Y$_m$  compounds under study,
   obtained from experimental values of the heat of formation and of
   the vibrational frequencies of these molecules
   \cite{Schlegel:84,Ignacio:90,Su:93}.
For the compounds for which no such experimental data were available,
   the theoretical values of these quantities were used to calculate
   the Schr\"odinger energy. 
Columns 3--6 of Tab.~\ref{tab:theo} report the total energy obtained
   at the HF, MP2, MP3 and MP4SDTQ4 (MP4) levels.
In the last column, the BOCE energy is reported.
For the sake of simplicity, Tab.~\ref{tab:theo} does not include the
   molecules employed in the BOCE treatment to calculate the
   bond-order parameter $a_{AB}$.
Tab.~\ref{tab:exptheo} reports the experimental and the calculated
   correlation energies for the same molecules listed in
   Tab.~\ref{tab:theo}.

From Tab.~\ref{tab:theo}, it is evident that using the \emph{ab
   initio} methods, the successive corrections to the molecular energy
   are very small. 
The differences between the MP2 and MP3 results range in general
   between $0.02-0.04$~a.u., while the difference between the next
   successive corrections, MP3 and MP4, is even smaller
   ($0.01-0.02$~a.u.).
For some molecules (SiHF$_3$, SiHF$_2$, SiF$_2$), the total energy
   increases on passing from the MP2 to the MP3 correction.
This trend is better evidenced by the correlation energies in
   Tab.~\ref{tab:exptheo}.
There, the change in the values of $E_c$ ranges between 0.02 and
   0.05~a.u., on going from the MP2 to the MP4 correction.

The very good agreement of the calculated molecular energy using the
   BOCE technique with the Schr\"odinger value appears clearly from
   the values in Tabs.~\ref{tab:theo} and \ref{tab:exptheo}.
The highly accurate values of the correlation energy obtained from
   BOCE for all molecules of the series are confirmed from the
   calculated percentage error with respect to the experimental
   correlation energy, ranging within $0.002-4.06$~\%.
On the contrary, \emph{ab initio} correlation energies yield very high
   percentage errors.
In Tab.~\ref{tab:theo}, for some molecules, the calculated correlation
   energy with the BOCE method gives a value higher than the
   experimental one, and the corresponding molecular energy is lower
   than the Schr\"odinger energy.
The absolute percentage error for these molecules has a value in the
   range between 0.002~\% and 0.4~\%.
However, it is important to add some comments here.
The first one is that the Schr\"odinger energy is obtained from
   experimental data and any experimental measurement is subject to an
   error that, in general, is about $\pm 5$~\%.
The second comment, pertaining to the compounds corresponding to a
   molecular energy lower than the Schr\"odinger energy, is that one
   needs to note that in closed shell systems, such as those
   considered here, this happens only in two cases, where we have used
   some theoretical values (vibrational frequencies, or molecular heat
   of formation, or both) in order to determine the Schr\"odinger
   molecular energy.

\begin{figure}[t]
\centering
\includegraphics[height=0.9\columnwidth,angle=-90]{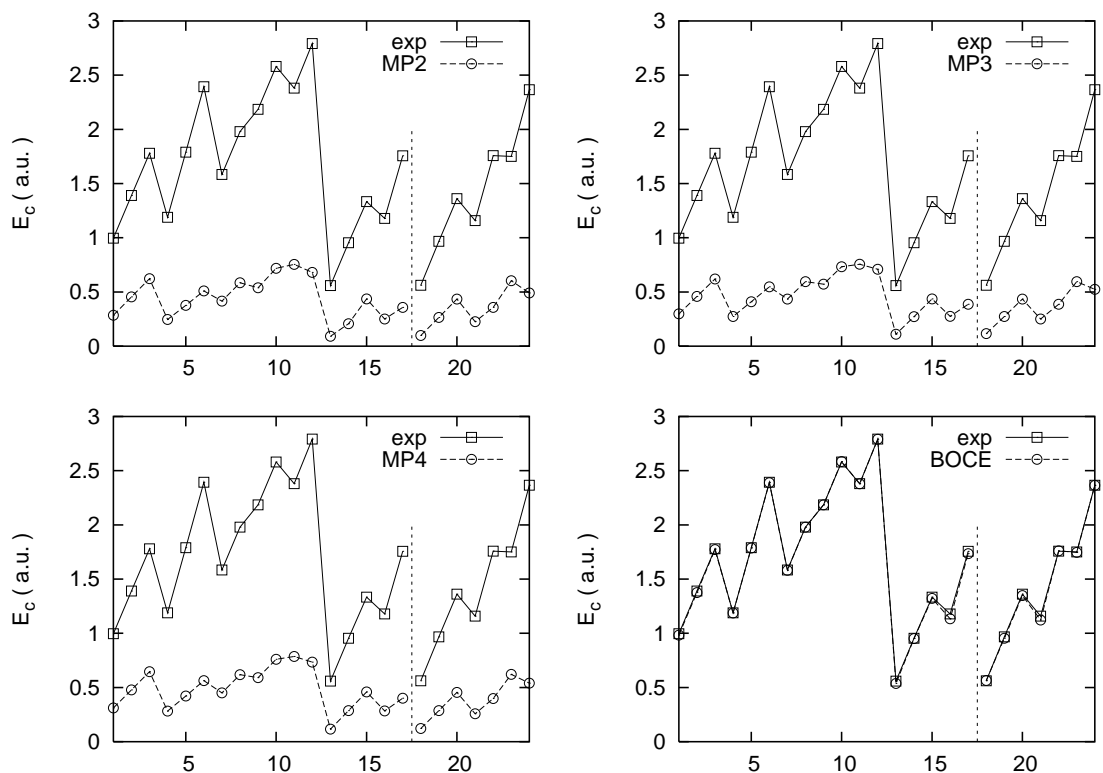}
\caption{Correlation energies $E_c$ (in a.u.) of closed and
   open shell silicon compounds, for the 24 molecules in
   Tab.~\ref{tab:exptheo}. 
The abscissa is the row index in Tab.~\ref{tab:exptheo}.
The vertical dashed line separates closed from open shell molecules.
In all insets, open squares refer to experimental values, while open
   circles refer to calculated values.}
\label{fig:exptheo}
\end{figure}

\begin{figure}[t]
\centering
\includegraphics[height=0.9\columnwidth,angle=-90]{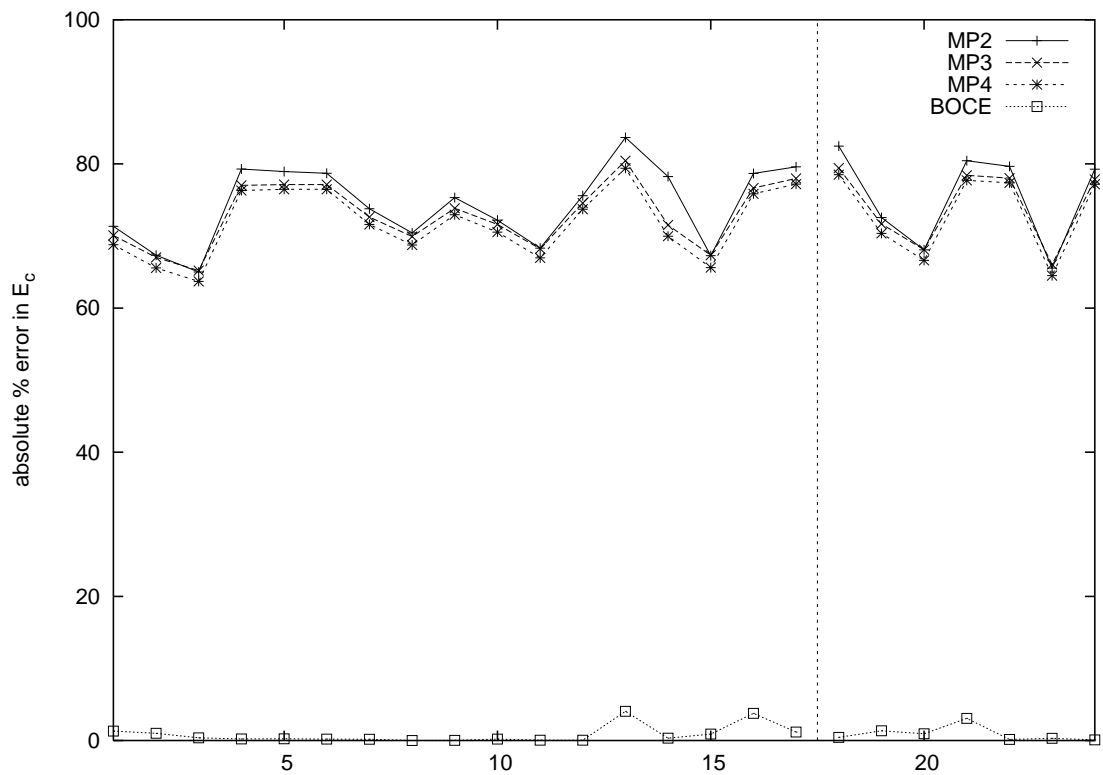}
\caption{Absolute percentage errors, $100|(E_c^{\mathrm{exp}} -
   E_c^{\mathrm{theo}})/E_c^{\mathrm{exp}} |$, of the theoretical correlation
   energies with respect to the experimental correlation
   energy in Tab.~\ref{tab:exptheo} and Fig.~\ref{fig:exptheo}.
The abscissa is the row index in Tab.~\ref{tab:exptheo}.
The vertical dashed line separates closed from open shell molecules.
}
\label{fig:percentage}
\end{figure}

The results of Tab.~\ref{tab:exptheo} are shown in
   Fig.~\ref{fig:exptheo}.
From this figure it can be seen that the trends of the calculated
   correlation energies using \emph{ab initio} methods are close to
   the experimental behaviour, but the values are on average
   1.5~a.u. higher than the experimental values.
Moreover, the increase in the calculated correlation energy on passing
   from MP2 to MP4 is negligible and, taking into account that the
   computational effort increases considerably from MP2 to MP4, we can
   conclude that it is not necessary to make corrections at higher
   order than MP2.
This is confirmed by Fig.~\ref{fig:percentage}, where the absolute
   percentage error between the experimental and theoretical
   correlation energies is almost constant for each compound on
   passing from MP2 to MP4.

The extent of the agreement between the experimental and the
   calculated BOCE correlation energies is shown in
   Fig.~\ref{fig:exptheo} (last panel), where the experimental and the
   BOCE curves are almost superimposed and hardly distinguishable.
Consequently, as shown in Fig.~\ref{fig:percentage}, the percentage
   error between the experimental and the calculated BOCE correlation
   energies is nearly constant, and varies between 0.002~\% and 4.0~\%.

\subsection{Ionization potentials (IP)}
\label{ssec:ip}

\begin{figure}[t]
\centering
\includegraphics[width=0.5\columnwidth]{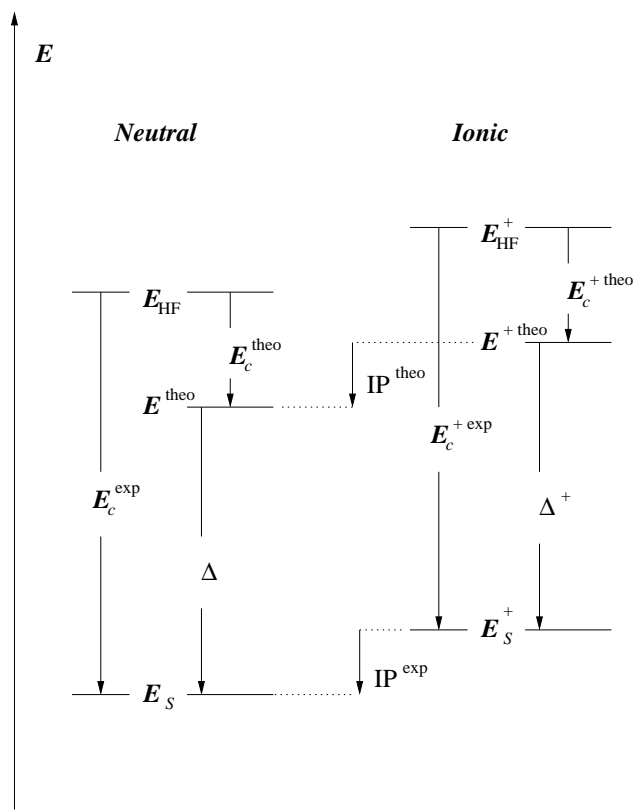}
\caption{Schematic determination of the experimental and theoretical
   correlation energies for both neutral and ionic molecules.}
\label{fig:schema}
\end{figure}

We have also calculated the ionization potentials (IP) of some
   SiX$_n$Y$_m$ compounds, using both \emph{ab initio} and BOCE
   methods.
These results have then been compared with the experimental values.
Within the so-called $\Delta$SCF procedure, the theoretical values of
   IP (in eV) are defined as the difference between the total energies
   of the neutral and the ionized molecule, \emph{i.e.}
\begin{equation}
\mathrm{IP} = E - E^+ 
\end{equation}
(see also Fig.~\ref{fig:schema}).
Table~\ref{tab:delta} reports the experimental and calculated IP for
   the molecules under study.
In these calculations, the geometries of both the neutral and the ionic
   species have been optimized.

\begin{figure}[t]
\centering
\includegraphics[height=0.9\columnwidth,angle=-90]{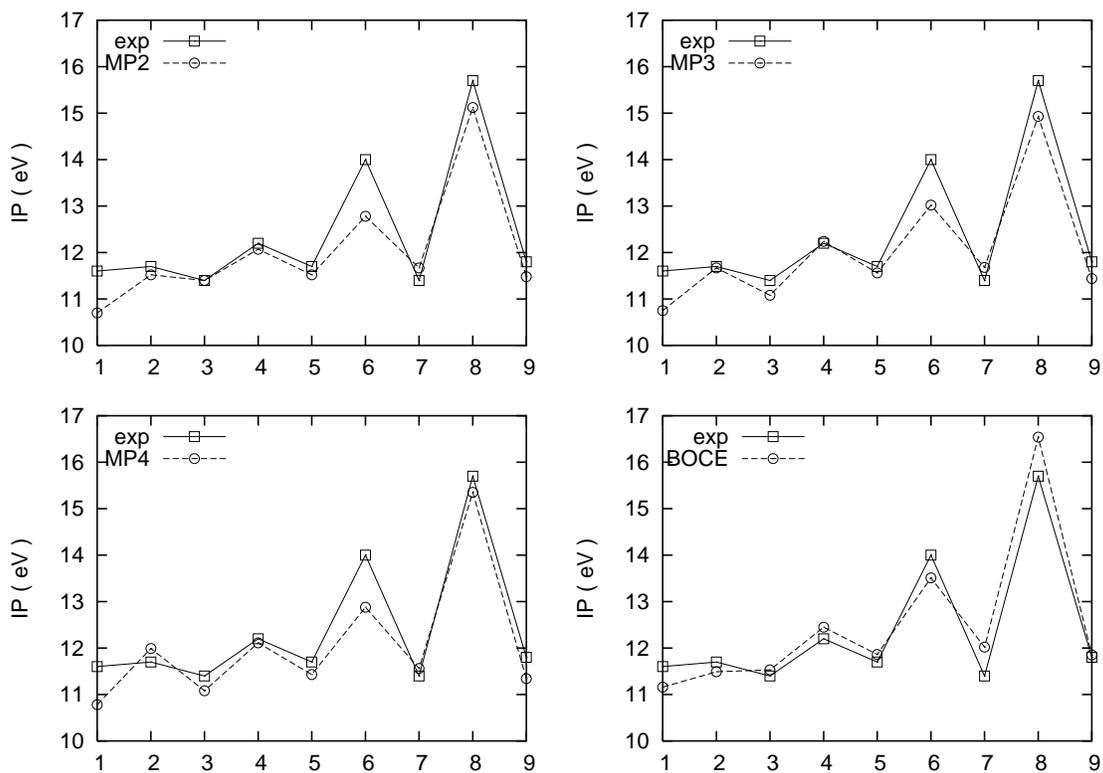}
\caption{Ionization potentials IP (in eV) for the 9 Si-containing
   molecules in Tab.~\ref{tab:delta}. 
The abscissa is the row index in Tab.~\ref{tab:delta}.
Open squares refer to experimental values, while open circles refer to
   calculated values. 
}
\label{fig:delta}
\end{figure}

In the second column of Tab.~\ref{tab:delta} we record the IP values
   obtained as the difference $E-E^+$, both calculated at HF level
   (without correlation), while in columns 5, 7, and 9 this difference
   refers to the calculated values at the MP2, MP3, and MP4 levels,
   respectively.
Finally, column~11 reports the calculated IP values using the BOCE
   method.
Fig.~\ref{fig:delta} displays the data in Tab.~\ref{tab:delta}.

\begin{figure}[t]
\centering
\includegraphics[height=0.9\columnwidth,angle=-90]{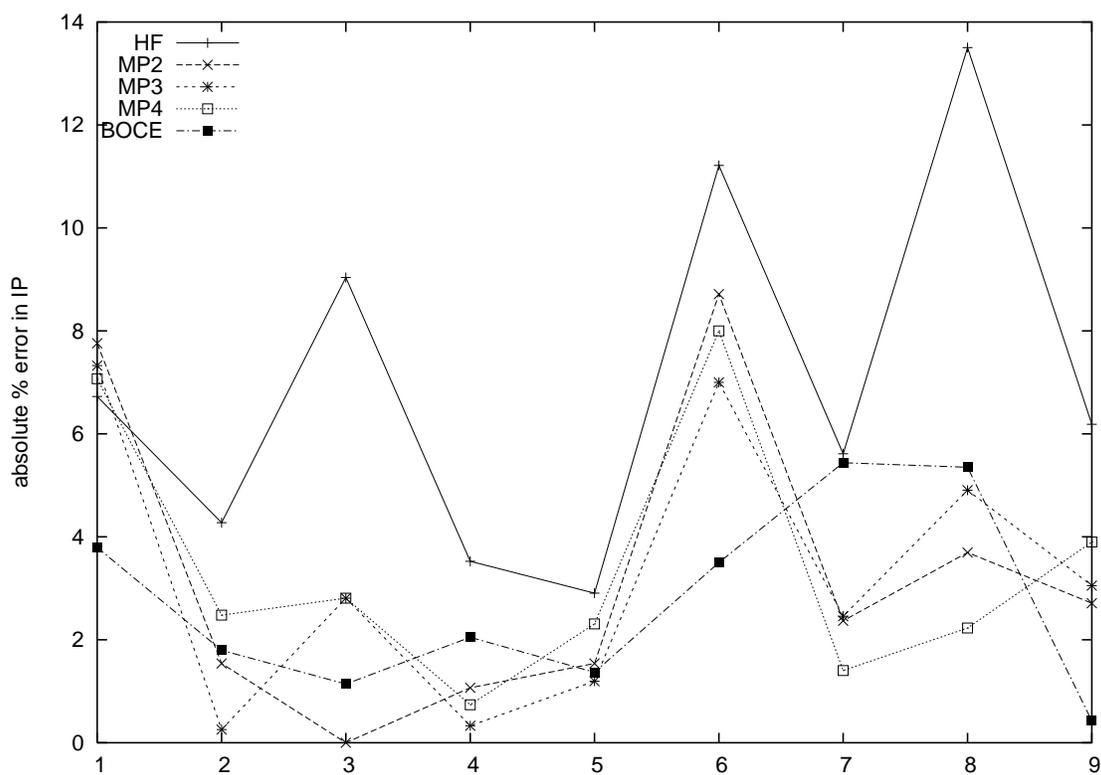}
\caption{Absolute percentage errors,
   $100|(\mathrm{IP}^{\mathrm{exp}} -
   \mathrm{IP}^{\mathrm{theo}})/\mathrm{IP}^{\mathrm{exp}} |$, of the
   theoretical ionization potentials with respect to the experimental
   values in Tab.~\ref{tab:delta} and Fig.~\ref{fig:delta}.
The abscissa is the row index in Tab.~\ref{tab:delta}.
}
\label{fig:ip}
\end{figure}

From Tab.~\ref{tab:delta} it is clear that the MP corrections,
   contrary to the case of the neutral compounds, are important in the
   IP calculation.
In fact, in all compounds of this series (with the exception of
   SiH$_4$), the absolute percentage error of the IP at the HF level
   is very high ($\sim 7$~\%; see Fig.~\ref{fig:ip}), and the
   introduction of correlation effects at MP2 level reduces this error
   to 3.26~\%, on the average.
Then the introduction of correlation in the estimate of IP yields a
   calculated value closer to the experimental one.
The trend of the next corrections, both at the MP3 and at the MP4
   levels, is not constant.
For some compounds of this series, SiH$_4$, SiH$_2$F$_2$, SiHCl$_3$,
   and SiF$_4$, the percentage error decreases on passing from MP2 to
   MP4.
On the contrary, for other compounds (SiCl$_4$, SiH$_3$Cl) the error
   increases, while for SiHF$_3$ and SiH$_2$Cl$_2$ the best estimate
   of IP is obtained at the MP3 level.
The average absolute percentage errors for the MP2, MP3, and MP4
   corrections are 3.26~\%, 3.26~\%, and 3.44~\%, respectively, thus
   showing that all these low order MP corrections give
   almost equivalent approximations of the IP.
Coming now to the BOCE approach, from the absolute percentage errors
   plotted in Fig.~\ref{fig:ip} one may conclude that this method
   yields more accurate estimates of the IP, on the average, than the
   \emph{ab initio} methods, although the MP methods occasionally
   produce lower absolute percentage errors in IP than the BOCE
   method.

A final consideration concerns the calculated differences, $\Delta$,
   between the experimental and calculated IP, reported in
   Tab.~\ref{tab:delta}.
Fig.~\ref{fig:schema} schematically defines the experimental and
   theoretical correlation energies for both neutral and ionic
   molecules.
According to Fig.~\ref{fig:schema}, we may then deduce that
\begin{equation}
\mathrm{IP}^{\mathrm{exp}} - \mathrm{IP}^{\mathrm{theo}} =
   (E_c^{\mathrm{exp}} - E_c^{\mathrm{theo}} ) -
   (E_c^{+\,\mathrm{exp}} - E_c^{+\,\mathrm{theo}} ) = \Delta -
   \Delta^+ ,
\end{equation}
where a superscript $+$ refers to the same quantity in the ionic
   molecule.
In Tab.~\ref{tab:delta} we record the experimental and theoretical
   ionization potentials (IP), along with their difference, $\Delta$.
The fact that the average error in the calculation of the IP is not
   very different within the MP and BOCE methods, is an indication
   that there are large cancellations between the correlation energies
   of the neutral and the ionic molecules.
In turn, this means that within the BOCE approach also the correlation
   energy of the ions is well approximated by using the same bond
   parameters $a_{AB}$, as derived for the same model molecules. 

\section{Summary and concluding remarks}
\label{sec:summary}

The biotechnological possibilities of silicon-containing molecules in
   substrate layers, plus much current interest in understanding the
   quite different reaction mechanism of silicon compounds in
   comparison with carbon compounds, has motivated us to extend our
   earlier work \cite{Grassi:96} on the latter class to the case of
   Si. 

What seemed to us important in this different series of small
   molecules was to assess the utility of our earlier semi-empirical
   use of molecular bond-order to estimate electron correlation
   energies in some members of C-containing molecules when applied to
   SiX$_n$Y$_m$, where X, Y = H, F, Cl, and $n+m = 4$, 3, 2.
In this process, we have found it valuable to compare and contrast our
   bond-order approach with some ab initio results we have obtained
   using low-order M\o{}ller-Plesset perturbation theory.
While, by these two approaches, we record quite a number of useful
   results for molecular correlation energies of Si-containing
   molecules, we wish especially to stress that the semi-empirical
   bond-order approach proposed in \cite{Grassi:96} and applied there
   to C-containing molecules continues to be valuable for molecules of
   the class  SiX$_n$Y$_m$.
It seems particularly useful when ionicity plays a role, whereas the
   low-order M\o{}ller-Plesset approach comes into its own for
   essentially neutral members of this series.

In conclusion, we emphasize the essence of the present approach which
   is to yield an impressive empirical correlation for obtaining bond
   additivity corrections to the energy based on bond orders
   calculated from a L\"owdin population analysis.
In order to present rather extensive results for a series of
   Si-containing molecules, we have, in the present work, accepted the
   limitations of the small basis set used.
However, should it prove instructive for future purposes, it is a
   straightforward matter to extend the basis, even if somewhat
   time-consuming.
Plainly, then, some quantitative changes will occur and also some
   specific technical points can be examined such as the numerical
   modifications in the L\"owdin population analyses, and the
   parameters obtained from them, due to changes in the basis set.
Another matter deserving attention if eventually larger basis sets are
   used concerns specifically the Si-F bond, which present indications
   suggest is somewhat problematic in the calculations, or experiment,
   or perhaps both.

We are encouraged by the results of the present investigation to
   attempt a pilot study of the relevance of our bond-order
   considerations to the still more difficult area of the kinetics and
   reactivity of small Si-containing molecules.
We hope to report on this area, approached via our semi-empirical
   procedures, at a later date.

\appendix

\section{Calculation of the bond-order parameter $a_{\mathrm{SiH}}$}
\label{app:bond}

In this Appendix we describe the procedure employed to estimate the
   bond parameter $a_{AB}$ in Eq.~(\ref{eq:Loewdin}), when
   $\mathrm{AB}=\mathrm{SiH}$.
According to Eqs.~(\ref{eq:binding}) and (\ref{eq:Loewdin}), the
   experimental correlation energy of the SiH$_4$ molecule can be
   written as
\begin{equation}
E_c^{\mathrm{exp}} (\mathrm{SiH_4} ) = E_c^{\mathrm{exp}}
   (\mathrm{Si}_{\mbox{\scriptsize in molecule}} )
+ \sum_{j=1}^4 E_c^{\mathrm{exp}} (\mathrm{H}^{(j)}_{\mbox{\scriptsize in
   molecule}} )
+ a_{\mathrm{SiH}} \sum_{i=1}^4 P_{\mathrm{SiH}}^{(i)} 
+ a_{\mathrm{HH}} \sum_{i=1}^6 P_{\mathrm{HH}}^{(i)} ,
\label{eq:SiH4}
\end{equation}
where $P_{\mathrm{SiH}}$ and $P_{\mathrm{HH}}$ and the
   bond-orders for each Si--H and H--H bonds in SiH$_4$, respectively.
From the H$_2$ molecule one immediately obtains the bond parameter for
   the H--H bond as
\begin{equation}
a_{\mathrm{HH}} = \frac{E_c^{\mathrm{exp}} (\mathrm{H_2} ) -2E_c^{\mathrm{exp}}
   (\mathrm{H}_{\mbox{\scriptsize in molecule}} )}{P_{\mathrm{HH}}} .
\end{equation}
From Eq.~(\ref{eq:Zn}), with $Z=n=1$, one has $E_c^{\mathrm{exp}}
   (\mathrm{H}_{\mbox{\scriptsize in molecule}} )=0$.
Since in the hydrogen molecule $P_{\mathrm{HH}} = 1$, one finds
\begin{equation}
a_{\mathrm{HH}} = E_c^{\mathrm{exp}} (\mathrm{H_2} ) =
   4.28\cdot10^{-2} .
\end{equation}
In Tab.~\ref{tab:bom} we report the symmetric matrix of the bond-order
   for SiH$_4$.
The diagonal elements are the sum over the off-diagonal elements, and
   represent the total electrons involved for each atom in the bonds.

Using the experimental value $E_c^{\mathrm{exp}} (\mathrm{SiH_4} ) =
   2.1742$~a.u., and employing the results of Tables~\ref{tab:exptheo}
   and \ref{tab:bom}, from Eq.~(\ref{eq:SiH4}) we eventually find
\begin{equation}
a_{\mathrm{SiH}} = 6.238\cdot 10^{-2} ,
\end{equation}
as quoted by Tab.~\ref{tab:bond}.

\bibliographystyle{molphys}
\bibliography{a,b,c,d,e,f,g,h,i,j,k,l,m,n,o,p,q,r,s,t,u,v,w,x,y,z,zzproceedings,Angilella}

\newpage

\begin{table}[t]
\centering
\begin{tabular}{|c|c|r@{E}lr@{E}lr@{E}l|}
\hline
Atom & $S$ & \multicolumn{2}{c}{$E_\HF$} & \multicolumn{2}{c}{$E_S$} &
   \multicolumn{2}{c|}{$E_c^{\mathrm{exp}}$} \\
\hline
Si & 3 & $-2.888318$ & $+02$ & $-2.893116$ & $+02$ & $4.798144$ & $-01$ \\
H & 2 & $-4.982329$ & $-01$ & $-4.993000$ & $-01$ & $1.067100$ & $-03$ \\
F & 2 & $-9.936496$ & $+01$ & $-9.972500$ & $+01$ & $3.600440$ & $-01$ \\
Cl & 2 & $-4.594480$ & $+02$ & $-4.600301$ & $+02$ & $5.821385$ & $-01$ \\
\hline
\end{tabular}
\caption{Hartree-Fock energy, $E_\HF$, estimated with the
   6-31G$^{\ast\ast}$ basis set, Schr\"odinger energy, $E_S$,
   and experimental correlation energy, $E_c^{\mathrm{exp}}$, for Si, H, F and
   Cl atoms.
The second column lists the values of the spin multiplicity, $S$.
All values are in a.u.}
\label{tab:HF}
\end{table}

\begin{table}[c]
\centering
\begin{tabular}{|c|c|l@{E}r|}
\hline
$A$--$B$ & $AB_n$ & \multicolumn{2}{c|}{$a_{AB}$} \\
\hline
H--H & H$_2$ & 4.287390 & $-02$ \\
Si--H & SiH$_4$ & 6.237950 & $-02$ \\
F--F & F$_2$ & 1.674987 & $-01$ \\
Cl--Cl & Cl$_2$ & 1.314397 & $-01$ \\
H--F & HF & 1.137677 & $-01$ \\
H--Cl & HCl & 7.700011 & $-02$ \\
Si--F & SiF$_4$ & 1.273342 & $-01$ \\
Si--Cl & SiCl$_4$ & 1.024566 & $-01$ \\
F--Cl & FCl & 1.586152 & $-01$ \\
\hline
\end{tabular}
\caption{L\"owdin bond parameters $a_{AB}$ for $A$--$B$ bonds in
   several $AB_n$ model molecules.
See Appendix~\ref{app:bond} for the derivation of $a_{\mathrm{SiH}}$.
For the FCl molecule (last row), the experimental vibrational
   frequency being unavailable, a theoretical value has been used in
   the calculation of $E_S$.} 
\label{tab:bond}
\end{table}

\begin{table}[c]
\centering
\begin{small}
\begin{tabular}{|c|r@{.}lr@{.}lr@{.}lr@{.}lr@{.}lr@{.}l|}
\hline
Molecule & \multicolumn{2}{c}{$E_S$} & \multicolumn{2}{c}{$E_\HF$} &
   \multicolumn{2}{c}{MP2} & \multicolumn{2}{c}{MP3} &
   \multicolumn{2}{c}{MP4} & \multicolumn{2}{c|}{BOCE} \\
\hline
   SiH$_3$F & $-$391 & 149094 & $-$390 & 152840 & $-$390 & 438375 & $-$390 & 450934 &
   $-$390 & 463879 & $-$391 & 136150 \\
   SiH$_2$F$_2$ & $-$490 & 474680 & $-$489 & 084835 & $-$489 & 538902 & $-$489 & 543176 &
   $-$489 & 563139 & $-$490 & 460715 \\
   SiHF$_3$ & $-$589 & 799528 & $-$588 & 019851 & $-$588 & 642790 & $-$588 & 639082 &
   $-$588 & 665926 & $-$589 & 792771 \\
   SiH$_3$Cl & $-$751 & 377125 & $-$750 & 187745 & $-$750 & 434135 & $-$750 & 461123 &
   $-$750 & 469475 & $-$751 & 374545 \\
   SiH$_2$Cl$_2$ & $-$1210 & 934066 & $-$1209 & 143833 & $-$1209 & 520705 & $-$1209 & 553336
   & $-$1209 & 564921 & $-$1210 & 929943 \\
   SiHCl$_3$ & $-$1670 & 491322 & $-$1668 & 097366 & $-$1668 & 607261 & $-$1668 & 644964
   & $-$1668 & 660441 & $-$1670 & 486756 \\
SiH$_2$FCl$^{a,b}$ & $-$850 & 697705 & $-$849 & 113982 & $-$849 & 529345 & $-$849 & 547673 &
   $-$849 & 563687 & $-$850 & 694892 \\
SiHF$_2$Cl$^{a,b}$ & $-$950 & 024261 & $-$948 & 045333 & $-$948 & 630412 & $-$948 & 640346 &
   $-$948 & 663824 & $-$950 & 024303 \\
SiHFCl$_2$$^{a,b}$ & $-$1310 & 255828 & $-$1308 & 071123 & $-$1308 & 610000 & $-$1308 & 642327
   & $-$1308 & 662046 & $-$1310 & 255383 \\
SiF$_2$Cl$_2$$^{a,b}$ & $-$1409 & 578850 & $-$1406 & 998802 & $-$1407 & 716386 & $-$1407 & 731725
   & $-$1407 & 759092 & $-$1409 & 583792 \\
   SiF$_3$Cl$^a$ & $-$1049 & 355060 & $-$1046 & 974456 & $-$1047 & 728506 & $-$1047 & 730616
   & $-$1047 & 761047 & $-$1049 & 353836 \\
   SiFCl$_3^a$ & $-$1769 & 814012 & $-$1767 & 022899 & $-$1767 & 704115 & $-$1767 & 732799
   & $-$1767 & 756713 & $-$1769 & 813079 \\
   SiH$_2$ & $-$290 & 561953 & $-$290 & 002631 & $-$290 & 093976 & $-$290 & 111950 &
   $-$290 & 117816 & $-$290 & 539281 \\
   SiHF & $-$389 & 887310 & $-$388 & 933412 & $-$389 & 140967 & $-$389 & 205213 &
   $-$389 & 219847 & $-$389 & 884142 \\
   SiF$_2$ & $-$489 & 218209 & $-$487 & 884672 & $-$488 & 321054 & $-$488 & 320264 &
   $-$488 & 343220 & $-$489 & 206212 \\
   SiHCl$^b$ & $-$750 & 122200 & $-$748 & 9442 & $-$749 & 195368 & $-$749 & 219098 &
   $-$749 & 228569 & $-$750 & 077830 \\
   SiCl$_2$ & $-$1209 & 699563 & $-$1207 & 943683 & $-$1208 & 301785 & $-$1208 & 330378 &
   $-$1208 & 344114 & $-$1209 & 678803 \\
\hline\hline
   SiH$_3^a$ & $-$291 & 173027 & $-$290 & 610579 & $-$290 & 709229 & $-$290 & 726369 &
   $-$290 & 731504 & $-$291 & 175489 \\
   SiH$_2$F$^a$ & $-$390 & 493513 & $-$389 & 525900 & $-$389 & 791422 & $-$389 & 800054 &
   $-$389 & 812671 & $-$390 & 480470 \\
   SiHF$_2^a$ & $-$489 & 813614 & $-$488 & 452373 & $-$488 & 886854 & $-$488 & 886530 &
   $-$488 & 906703 & $-$489 & 800833 \\
SiH$_2$Cl$^{a,b}$ & $-$750 & 723808 & $-$749 & 566110 & $-$749 & 792607 & $-$749 & 815941 &
   $-$749 & 823949 & $-$750 & 688190 \\
SiHCl$_2^{a,b}$ & $-$1210 & 279079 & $-$1208 & 521554 & $-$1208 & 878868 & $-$1208 & 907693
   & $-$1208 & 919212 & $-$1210 & 281674 \\
   SiF$_3$ & $-$589 & 132214 & $-$587 & 381225 & $-$587 & 984340 & $-$587 & 975648 &
   $-$588 & 002862 & $-$589 & 126715 \\
   SiCl$_3^a$ & $-$1669 & 840418 & $-$1667 & 474563 & $-$1667 & 964920
   & $-$1667 & 998734 & $-$1668 & 014351 & $-$1669 & 838233 \\
\hline
\end{tabular}
\end{small}
\caption{Schr\"odinger and theoretical energies (in a.u.) for open and
   closed shell of SiX$_n$Y$_m$ compounds.
The upper table refers to relevant energies of closed-shell silicon
   compounds: especially the bond-order correlation energy.
The lower table refers to the energies of open-shell silicon
   compounds.
{\sl \underline{Notes}:}
(a) The Schr\"odinger energies have been calculated using theoretical
   values of the vibrational frequencies.
(b) The Schr\"odinger energies have been calculated using theoretical
   values of the molecular heat of formation.
}
\label{tab:theo}
\end{table}

\begin{table}[t]
\centering
\begin{small}
\begin{tabular}{|c|r@{.}lr@{.}lr@{.}lr@{.}lr@{.}l|}
\hline
Molecule & \multicolumn{2}{c}{$E_c^{\mathrm{exp}}$} &
   \multicolumn{2}{c}{MP2} & 
   \multicolumn{2}{c}{MP3} & 
   \multicolumn{2}{c}{MP4} & 
   \multicolumn{2}{c|}{BOCE} \\
\hline
SiH$_3$F & 0 & 996255 & 0 & 285535 & 0 & 298094 & 0 & 310939 & 0 & 983310 \\
SiH$_2$F$_2$ & 1 & 389846 & 0 & 454068 & 0 & 458341 & 0 & 478385 & 1 & 375880 \\
SiHF$_3$ & 1 & 779678 & 0 & 622939 & 0 & 619231 & 0 & 646075 & 1 & 772920 \\
SiH$_3$Cl & 1 & 189380 & 0 & 246389 & 0 & 273378 & 0 & 281730 & 1 & 186800 \\
SiH$_2$Cl$_2$ & 1 & 790233 & 0 & 376872 & 0 & 409502 & 0 & 421088 & 1 & 786110 \\
SiHCl$_3$ & 2 & 393956 & 0 & 509894 & 0 & 547597 & 0 & 563074 & 2 & 389390 \\
SiH$_2$FCl$^{a,b}$ & 1 & 583723 & 0 & 415363 & 0 & 433691 & 0 & 449706 & 1 & 580910 \\
SiHF$_2$Cl$^{a,b}$ & 1 & 978928 & 0 & 585079 & 0 & 595013 & 0 & 618492 & 1 & 978970 \\
SiHFCl$_2^{a,b}$ & 2 & 184705 & 0 & 538877 & 0 & 571204 & 0 & 590923 & 2 & 184260 \\
SiF$_2$Cl$_2^{a,b}$ & 2 & 580048 & 0 & 717584 & 0 & 732924 & 0 & 760291 & 2 & 584990 \\
SiF$_3$Cl$^a$ & 2 & 380604 & 0 & 754049 & 0 & 756160 & 0 & 786592 & 2 & 379380 \\
SiFCl$_3^a$ & 2 & 791113 & 0 & 681216 & 0 & 709900 & 0 & 733814 & 2 & 790180 \\
SiH$_2$ & 0 & 559322 & 0 & 091345 & 0 & 109319 & 0 & 115185 & 0 & 536650 \\
SiHF & 0 & 953899 & 0 & 207555 & 0 & 271801 & 0 & 286436 & 0 & 950730 \\
SiF$_2$ & 1 & 333537 & 0 & 436382 & 0 & 435591 & 0 & 458548 & 1 & 321540 \\
SiHCl$^b$ & 1 & 177958 & 0 & 251168 & 0 & 274897 & 0 & 284368 & 1 & 133630 \\
SiCl$_2$ & 1 & 755880 & 0 & 358102 & 0 & 386695 & 0 & 400432 & 1 & 735120 \\
\hline
\hline
SiH$_3^a$ & 0 & 562448 & 0 & 098650 & 0 & 115791 & 0 & 120925 & 0 & 564910 \\
SiH$_2$F$^a$ & 0 & 967613 & 0 & 265521 & 0 & 274154 & 0 & 286771 & 0 & 954570 \\
SiHF$_2^a$ & 1 & 361241 & 0 & 434481 & 0 & 434157 & 0 & 454329 & 1 & 348460 \\
SiH$_2$Cl$^{a,b}$ & 1 & 157698 & 0 & 226497 & 0 & 249830 & 0 & 257838 & 1 & 122080 \\
SiHCl$_2^{a,b}$ & 1 & 757525 & 0 & 357315 & 0 & 386139 & 0 & 397659 & 1 & 760120 \\
SiF$_3$ & 1 & 750989 & 0 & 603115 & 0 & 594423 & 0 & 621637 & 1 & 745490 \\
SiCl$_3^a$ & 2 & 365855 & 0 & 490357 & 0 & 524171 & 0 & 539788 & 2 & 363670 \\
\hline
\end{tabular}
\end{small}
\caption{Experimental and theoretical correlation energies (in a.u.)
   for open and closed shell of SiX$_n$Y$_m$ compounds.
The upper table refers to closed-shell, whereas the lower
   table refers to open-shell compounds.
{\sl \underline{Notes}:}
(a) The Schr\"odinger energies have been calculated using theoretical
   values of the vibrational frequencies. 
(b) The Schr\"odinger energies have been calculated using theoretical
   values of the molecular heat of formation.} 
\label{tab:exptheo}
\end{table}

\begin{table}[t]
\centering
\begin{small}
\begin{tabular}{|c|r@{.}l|r@{.}lr@{.}l|r@{.}lr@{.}l|r@{.}lr@{.}l|r@{.}lr@{.}l|r@{.}lr@{.}l|}
\hline
Molecule & \multicolumn{2}{c|}{IP$^{\mathrm{exp}}$} & 
\multicolumn{2}{c}{HF} &
\multicolumn{2}{c|}{$\Delta$} &
\multicolumn{2}{c}{MP2} &
\multicolumn{2}{c|}{$\Delta$} &
\multicolumn{2}{c}{MP3} &
\multicolumn{2}{c|}{$\Delta$} &
\multicolumn{2}{c}{MP4} &
\multicolumn{2}{c|}{$\Delta$} &
\multicolumn{2}{c}{BOCE} &
\multicolumn{2}{c|}{$\Delta$} \\
\hline
SiH$_4$ & 11 & 60 & 12 & 38 & $-$0 & 78 & 10 & 70 & 0 & 90 & 10 & 75 & 0 & 85 & 10 & 78 &
   0 & 82 & 11 & 16 & 0 & 44 \\
SiH$_3$F & 11 & 70 & 12 & 20 & $-$0 & 5 & 11 & 52 & 0 & 18 & 11 & 67 & 0 & 03 & 11 & 99 &
   $-$0 & 29 & 11 & 49 & 0 & 21 \\
SiH$_3$Cl & 11 & 40 & 10 & 37 & 1 & 03 & 11 & 40 & 0 & 00 & 11 & 08 & 0 & 32 & 11 & 08 &
   0 & 32 & 11 & 53 & $-$0 & 13 \\
SiH$_2$F$_2$  & 12 & 20 & 11 & 77 & 0 & 43 & 12 & 07 & 0 & 13 & 12 & 24 & $-$0 & 04 & 12 & 11
   & 0 & 09 & 12 & 45 & $-$0 & 25 \\
SiH$_2$Cl$_2$ & 11 & 70 & 11 & 36 & 0 & 34 & 11 & 52 & 0 & 18 & 11 & 56 & 0 & 14 & 11 & 43 &
   0 & 27 & 11 & 86 & $-$0 & 16 \\
SiHF$_3$ & 14 & 00 & 12 & 43 & 1 & 57 & 12 & 78 & 1 & 22 & 13 & 02 & 0 & 98 & 12 & 88 &
   1 & 12 & 13 & 51 & 0 & 49 \\
SiHCl$_3$ & 11 & 40 & 12 & 04 & $-$0 & 64 & 11 & 67 & $-$0 & 27 & 11 & 68 & $-$0 & 28 & 11 & 56
   & $-$0 & 16 & 12 & 02 & $-$0 & 62 \\
SiF$_4$ & 15 & 70 & 17 & 82 & $-$2 & 12 & 15 & 12 & 0 & 58 & 14 & 93 & 0 & 77 & 15 & 35 &
   0 & 35 & 16 & 54 & $-$0 & 84 \\
SiCl$_4$ & 11 & 80 & 12 & 53 & $-$0 & 73 & 11 & 48 & 0 & 32 & 11 & 44 & 0 & 36 & 11 & 34 &
   0 & 46 & 11 & 85 & $-$0 & 05 \\
\hline
\end{tabular}
\end{small}
\caption{Experimental and theoretical ionization potentials (in eV).
The $\Delta$ columns report the differences between the experimental
   and the theoretical values.}
\label{tab:delta}
\end{table}

\begin{table}[t]
\centering
\begin{tabular}{|c|ccccc|}
\cline{2-6} 
\multicolumn{1}{c|}{} & Si & H & H & H & H \\
\hline
Si & 3.970845 & & & & \\
H & 0.993248 & 1.015563 & & & \\
H & 0.993248 & 0.006901 & 1.015563 & & \\
H & 0.991655 & 0.007873 & 0.007873 & 1.014972 & \\
H & 0.992695 & 0.007541 & 0.007541 & 0.007571 & 1.015348 \\
\hline
\end{tabular}
\caption{Bond-order matrix of SiH$_4$.}
\label{tab:bom}
\end{table}

\newpage

\end{document}